\documentclass[conference]{IEEEtran}

\usepackage{multirow}
\usepackage{amsmath}
\usepackage{float}
\usepackage{subcaption}
\usepackage{graphicx}
\usepackage[utf8]{inputenc} 
\usepackage[T1]{fontenc}    
\usepackage{hyperref}       
\usepackage{url}            
\usepackage{booktabs}       
\usepackage{amsfonts}       
\usepackage{nicefrac}       
\usepackage{lipsum}

\title{Digital Low-Level RF control system for Accumulator Ring at Advanced Light Source Upgrade Project}

\author{
    \IEEEauthorblockN{
        Qiang Du\IEEEauthorrefmark{1},
        Shreeharshini Murthy,
        Michael Betz,
        Kevin Bender,
        Wayne Lewis,
        Najm Us Saqib, \\
        Sergio Paiagua,
        Lawrence Doolittle,
        Carlos Serrano,
        Benjamin Flugstad,
        Kenneth Baptiste
    }
    \IEEEauthorblockA{
        Lawrence Berkeley National Laboratory\\
        1 Cyclotron Rd, Berkeley, CA, 94720 USA\\
        Email: QDu@lbl.gov
    }
}

\begin{document}
\maketitle

\begin{abstract}
    Currently ALS is undergoing an upgrade to ALS-U to produce 100 times
    brighter soft X-ray light.  The LLRF system for Accumulator Ring (AR) is
    composed of two identical LLRF stations, for driving RF amplifiers. The
    closed loop RF amplitude and phase stability is measured as $< 0.1\%$ and $<
    0.1 ^\circ$ respectively, using the non-IQ digital down conversion together
    with analog up/down conversion, under a system-on-chip architecture.
    Realtime interlock system is implemented with $< 2 \mu$s latency, for
    machine protection against arc flash and unexpected RF power.  Control
    interfaces are developed to enable PLC-FPGA-EPICS communication to support
    operation, timing, cavity tuning, and interlock systems. The LLRF system
    handles alignment of buckets to swap beams between AR and Storage Ring
    by synchronous phase loop ramping between the two cavities. The system also
    includes an optimization routine to characterize the loop dynamics and
    determine optimal operating point using a built-in network analyzer feature.
    A cavity emulator of 31 kHz bandwidth is integrated with the LLRF system to
    validate the performance of the overall system being developed.
\end{abstract}

\section{Introduction}

The Advanced Light Source (ALS) at Lawrence Berkeley National Laboratory is a
U.S. Department of Energy's synchrotron light source user facility that is
operational since 1993.  With circumference of 196.8 m, the ALS Storage Ring
(SR) keeps electron beam current of 500 mA at 1.9 GeV under multi--bunch mode
user operation to deliver synchrotron X-rays to surrounding 40 experimental end
stations. ~\cite{du2019digital}

\begin{figure}[H]
    \centering
    \includegraphics[width=0.8\linewidth]{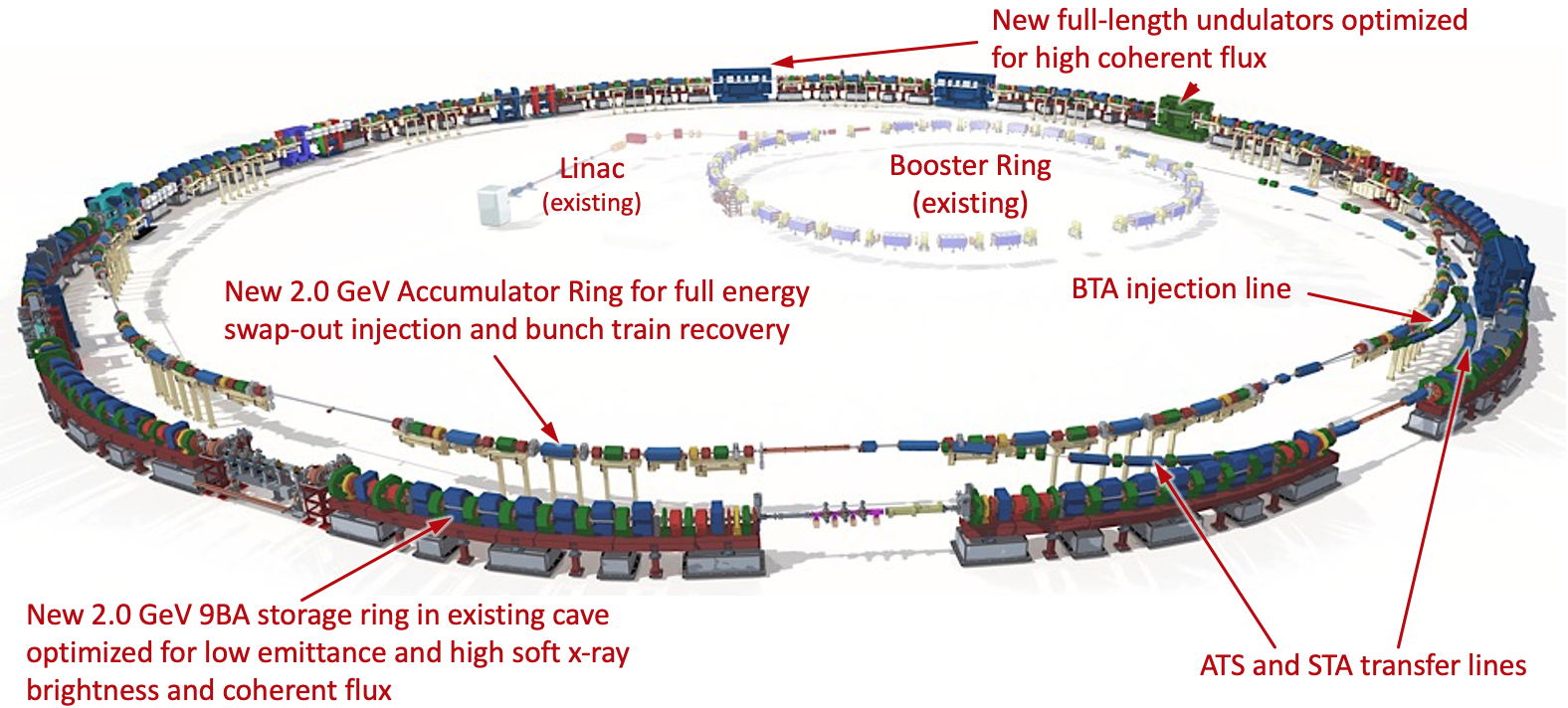}
    \includegraphics[width=0.8\linewidth]{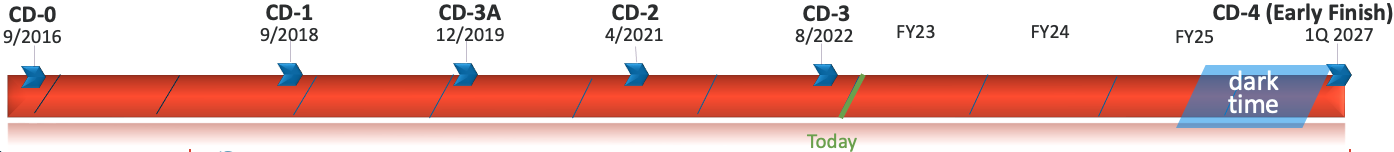}
    \caption{ALS Upgrade with both Accumulator and Storage Rings}
    \label{fig:alsu_rings}
\end{figure}

There is an ongoing ALS upgrade project (ALS-U), scheduled to upgrade towards 2
orders of magnitude increase in brightness and flux of 1keV soft X-rays at
diffraction limit.  The ALS-U project involves a new 2.0 GeV Storage Ring (SR)
in existing tunnel optimized for low emittance, and add a new 2.0 GeV
Accumulator Ring (AR) for full energy swap-out injection and bunch train
recovery, as shown in Fig.~\ref{fig:alsu_rings}.
The AR is a triple-bend-achromat (TBA) lattice, very similar to the current ALS SR lattice.
The RF system requirements for the TBA lattice can be found in
Table~\ref{table:rf_parameters}.
Two normal conducting RF cavities have been selected for AR.
Each RF cavity is driven by an identical chain from a high power solid-state amplifier (SSA) and
low-level RF (LLRF) control system, together with
personnel safety interlocks, equipment protection interlocks, etc.

\begin{table}[H]
    \centering
    \begin{tabular}{lcccc}
    \toprule
                & ALS SR & ALS-U SR & ALS-U AR & \\
    \midrule
        Cavity RF Frequency & 499.64    & 500.394     & 500.394 & MHz   \\
        Number of Cavities  & 2         & 2           & 2       &       \\
        $\frac{R}{Q}$ (ea)  & 4.9       & 4.9         & 3.4--3.5 & M$\Omega$\\
        Cavity voltage      & 671       & 300         & 500 & kV    \\
        $\beta$             & 2.9       & 10.6        & 1.18 &       \\
        Energy loss per trun& 329       & 347         & 270 & keV   \\
        BM Beam Power       & 141       & 125         & 13.3 & kW    \\
        ID Beam Power       & 42        & 35          &  & kW    \\
        3HC Beam Power      & 7.3       & 13.8        &  & kW    \\
        Parasitic Beam Power& 2.9 (est.)& 2.6 (est.)  & 0.2 & kW    \\
        Total Beam Power    & 192.9     & 176.4       & 13.5 & kW    \\
        Cavity Power(no beam)&46        & 9.2         & 36.0 & kW    \\
        Cavity Power(beam)  & 142.5     & 97.4        & 42.7 & kW    \\
        High Power Amplifier& 294.0     & 197.5       & 60 & kW    \\
    \bottomrule
    \end{tabular}
    \caption{ALS and ALS-U AR and SR RF parameters}
    \label{table:rf_parameters} 
\end{table}


\section{Hardware Design}
The AR LLRF system consists of two identical stations, each controlling one SSA and RF cavity,
but sharing communication interface with a central interlock PLC and both stations are coordinated By
common timing events for synchronous operations.
Each LLRF station consists of an analog frontend chassis and a digital chassis.
The analog chassis receives ALS-U master oscillator and generates local oscillator (LO) signal, 
and provides frequency conversions including 6 channels of down-conversion and 2 channels of up-conversion
between RF and IF frequencies, following the non-IQ down-conversion configuration:
\begin{align*}
    f_\text{MO} &= 500.39\,\text{MHz}\\
    f_\text{LO} &= \frac{11}{12} f_\text{MO} = 458.69\,\text{MHz}\\
    f_\text{IF} &= \frac{ 1}{12} f_\text{MO} = 41.69\,\text{MHz}\\
    f_\text{Sample} &=  \frac{1}{4} f_\text{LO} = 114.67\,\text{MHz}\\
    \frac{f_\text{IF}}{f_\text{Sample}} &= \frac{4}{11} = \theta \simeq 130.9^\circ
\end{align*}

\begin{figure}[H]
    \centering
    \includegraphics[width=0.9\linewidth, angle=90]{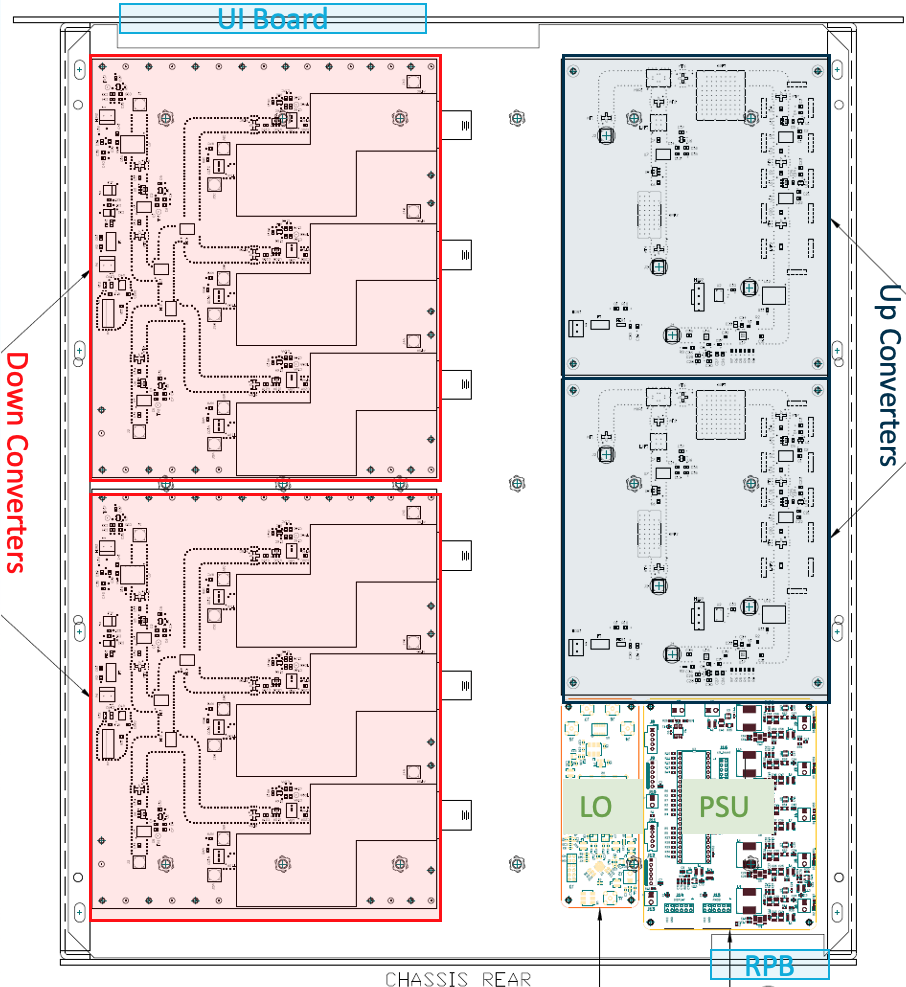}
    \caption{Analog frontend chassis: 6 down + 2 up conversion, LO generation}
    \label{fig:analog_chassis_assembly}
\end{figure}

The analog frontend chassis assembly is shown as in
Fig.~\ref{fig:analog_chassis_assembly}.  The up-down-conversion PCB boards are
existing designs from Jefferson Lab with 500 MHz center frequency, featured with
high channel isolation, programmable TX attenuation, RF switch and various
diagnostics and monitoring points.
The LO generation PCB board uses single side ban modulation to generate four $f_\text{LO}$
signals that feed into up-down-conversion boards, and the digital chassis for sampling clock.

There is a centralized power supply controller board that is responsible for
configuration of LO board, continuous monitoring and interlocking of the LO level, temperature, 
voltage and current consumptions, and provide an interactive local user interface through
an OLED display. Remote control is available through an RS485 port with Modbus RTU protocol.

\begin{figure}[H]
    \centering
    \includegraphics[width=0.9\linewidth]{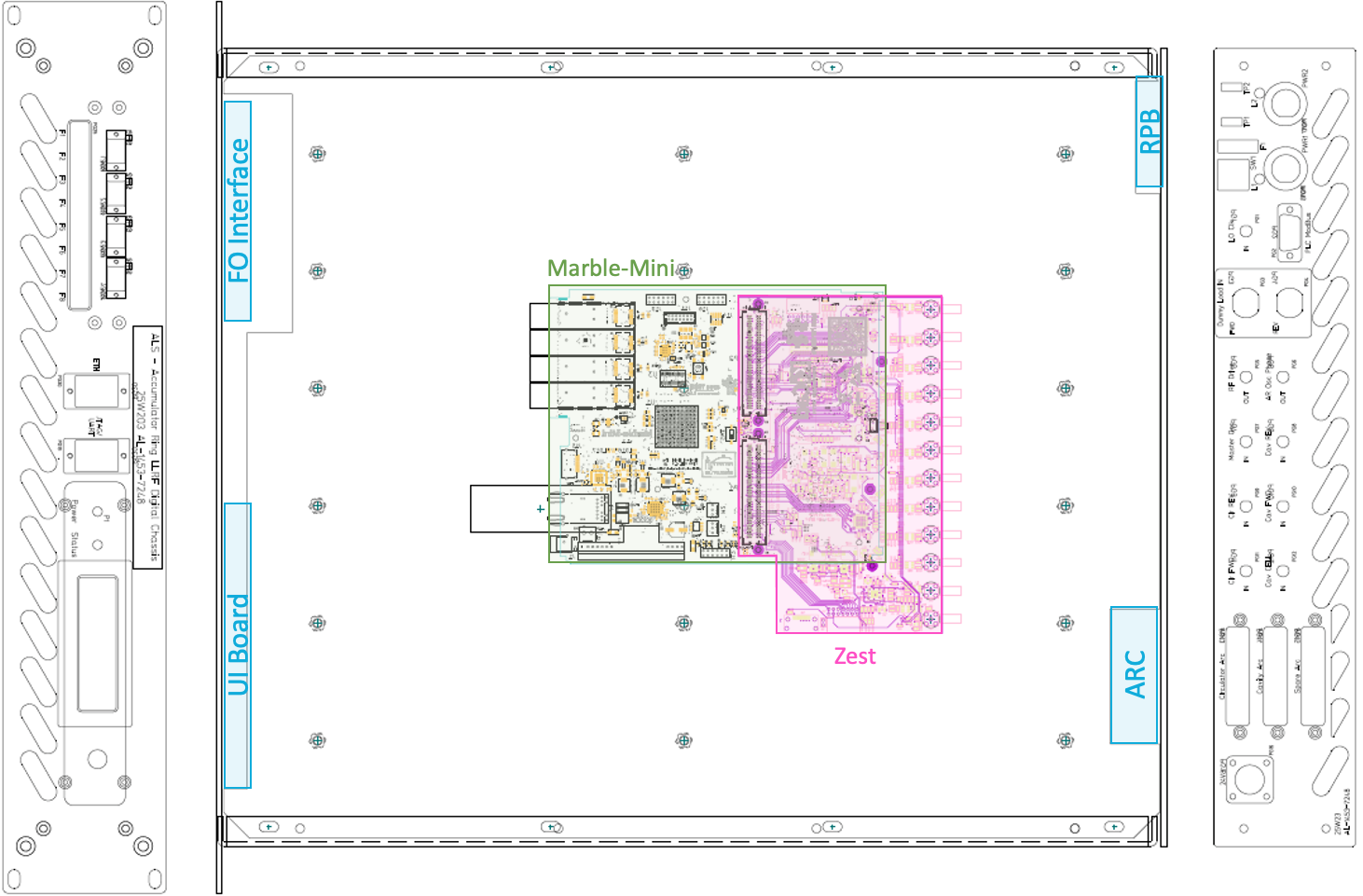}
    \caption{Digital chassis: Marble-Mini FPGA carrier and Zest digitizer}
    \label{fig:digital_chassis_assembly}
\end{figure}

The digital chassis consists a \href{https://github.com/BerkeleyLab/Marble-Mini}{Marble-Mini} ~\cite{lbl:marblemini}
FPGA carrier board and a \href{https://github.com/BerkeleyLab/Zest}{Zest} digitizer board ~\cite{lbl:zest},
as shown in Fig.~\ref{fig:digital_chassis_assembly}. 
Both boards are available through CERN Open Hardware License (OHL v1.2).
The $f_\text{LO}$ signal is feed into the clock input of Zest board for generation and 
distribution of $f_\text{Sample}$ to two 16-bit, four channel ADC with sampling rate up to 125 MHz (AD9653),
and one 16-bit, dual channel DAC with sampling rate up to 250 MHz (AD9781).
The Zest board is used in LCLS-II LLRF project with superior low noise and channel crosstalk performances.
The Marble family FPGA carrier boards are used in many other projects in ALS and PIP-II LLRF control.

For ALS-U AR LLRF control, the 6 down-converted IF signals are fed from the
analog frontend chassis for operation, including cavity probe, cavity forward
and reverse, and circulator load forward and reverse signals.  The 2 additional
ADCs in Zest board are used to directly sampling the test load signals for high
power RF system commissioning.

\section{Firmware design}

\begin{figure}[H]
    \includegraphics[width=0.9\linewidth]{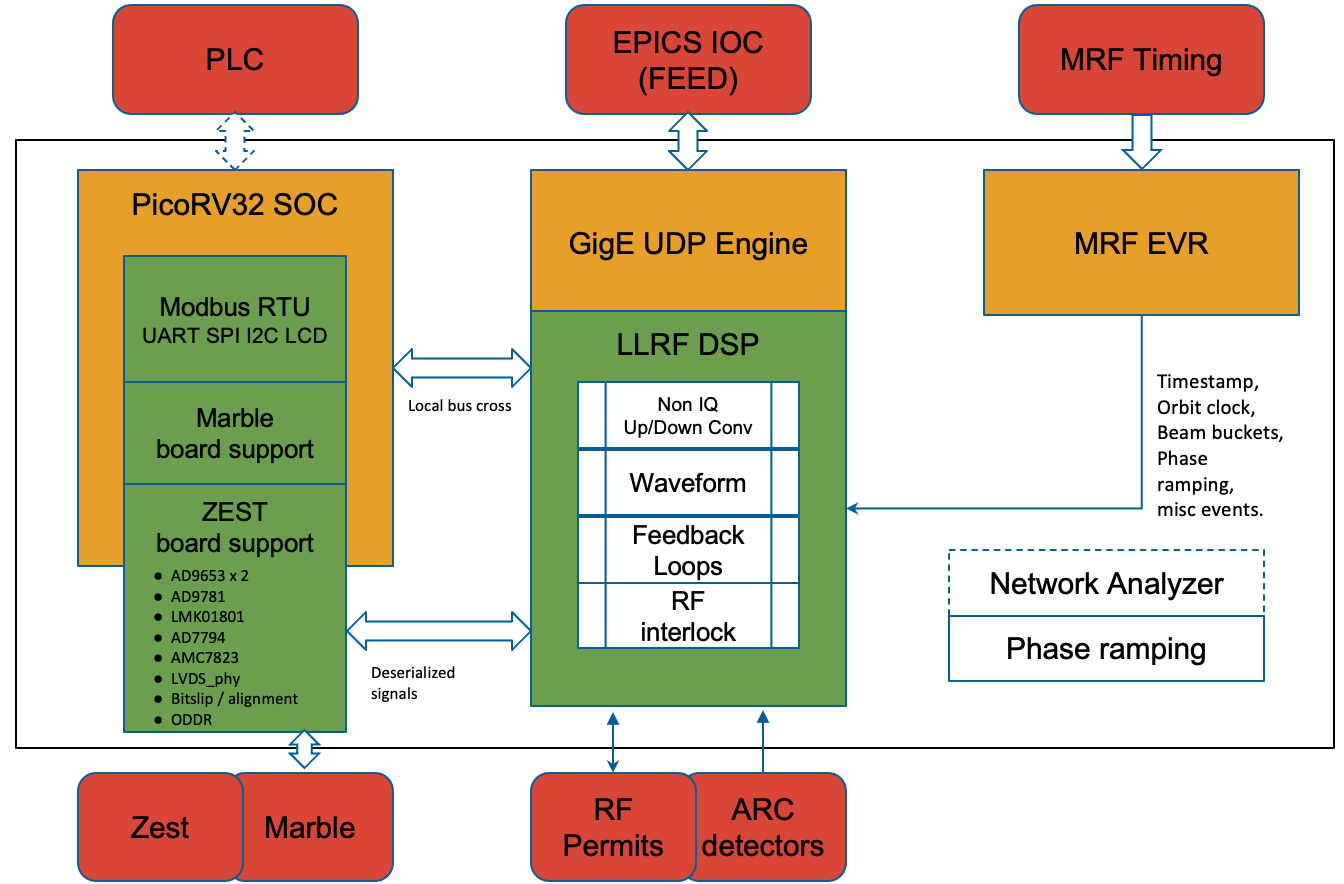}
    \caption{Digital Chassis Firmware architecture}
    \label{fig:firmware_arch}
\end{figure}

Fig.~\ref{fig:firmware_arch} shows the digital chassis firmware architecture, which is built
on top of BSD licensed open source LLRF library \href{https://github.com/BerkeleyLab/Bedrock}{Bedrock}~\cite{lbl:bedrock}.
We use an open source, size optimized RISC-V CPU \href{https://github.com/YosysHQ/picorv32}{PicoRV32}~\cite{clifford2015picorv32}
to handle system configuration, boot-time self checking, continuous status monitoring, and remote interfacing
with PLC systems via the RS485 / Modbus RTU port which is similar to the analog frontend chassis.
The footprint of the firmware is only 24 kB of RAM and 750--2000 LUTs.
At the same time, we use BerkeleyLab's realtime Ethernet fabric infrastructure \href{https://github.com/BerkeleyLab/Bedrock/tree/master/badger}{packet badger}
to provide a UDP communication as another local bus master, which has the priority to directly access DSP registers
for realtime EPICS applications, including configurable waveforms for any RF signals.

\begin{figure}[H]
    \includegraphics[width=0.9\linewidth]{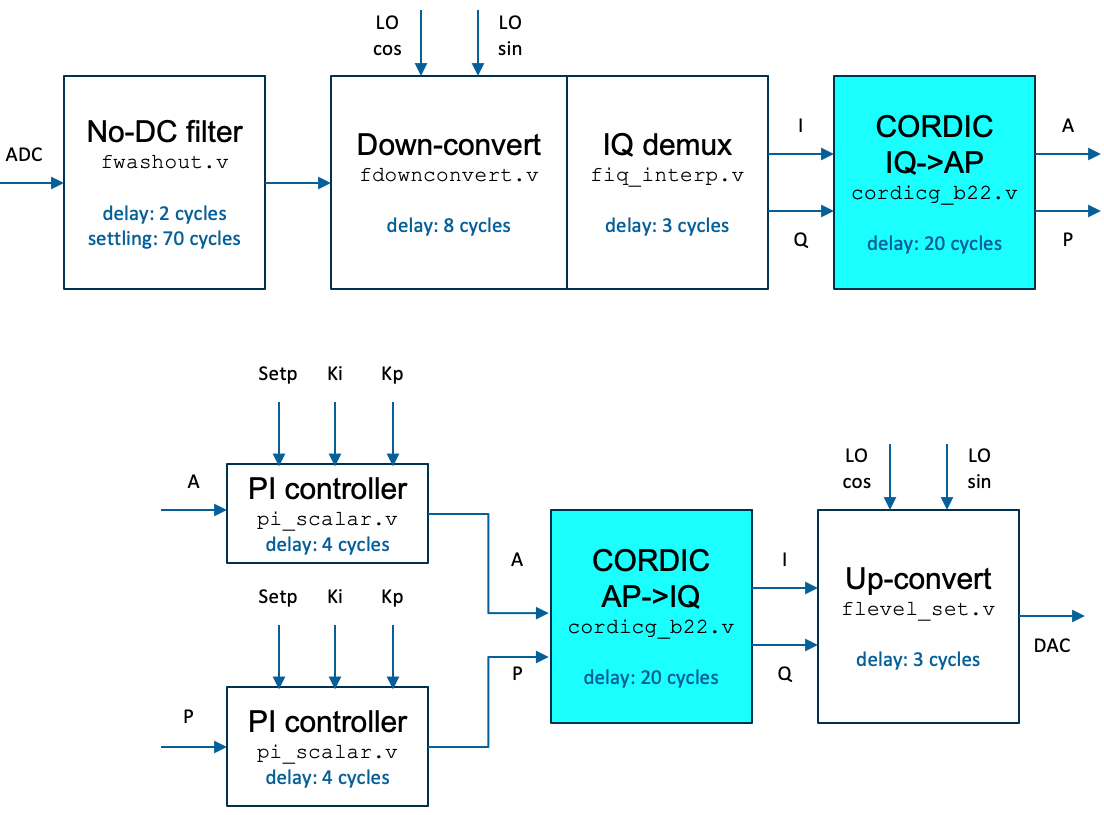}
    \caption{Feedback DSP core firmware implementation in fabric}
    \label{fig:firmware_dsp_core}
\end{figure}

The feedback DSP core is implemented as shown in Fig.~\ref{fig:firmware_dsp_core}.
A classical non-IQ digital down conversion algorithm~\cite{doolittle2006digital} is implemented
to digitally down-convert the IF signals and reconstruct their IQ values using a local direct digital synthesizer (DDS):
Given  $\frac{f_\text{IF}}{f_\text{Sample}} = \frac{4}{11} = \theta \simeq 130.9^\circ$,
\begin{align*}
    \begin{pmatrix}
        y_n \\
        y_{n+1}
    \end{pmatrix}
    &=
    \begin{pmatrix}
        \cos(n\theta)       & \sin(n\theta) \\
        \cos((n+1)\theta)   & \sin((n+1)\theta)
    \end{pmatrix}
    \begin{pmatrix}
        I \\
        Q
    \end{pmatrix} \\
    \begin{pmatrix}
        I \\
        Q
    \end{pmatrix}
    &= \frac{1}{\sin\theta}
    \begin{pmatrix}
        \sin((n+1)\theta)    & -\sin(n\theta) \\
        -\cos((n+1)\theta)   & \cos(n\theta)
    \end{pmatrix}
    \begin{pmatrix}
        y_n \\
        y_{n+1}
    \end{pmatrix}
\end{align*}

\begin{figure}
    \includegraphics[width=0.9\linewidth]{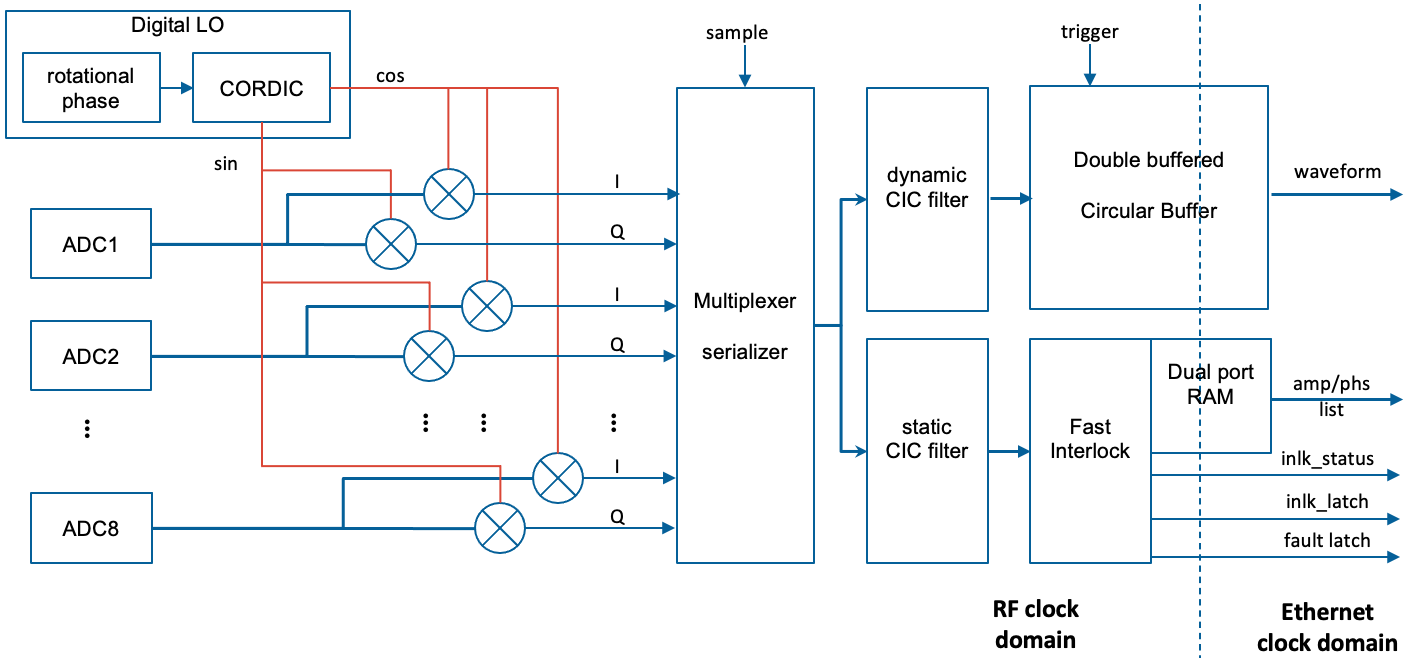}
    \includegraphics[width=0.9\linewidth]{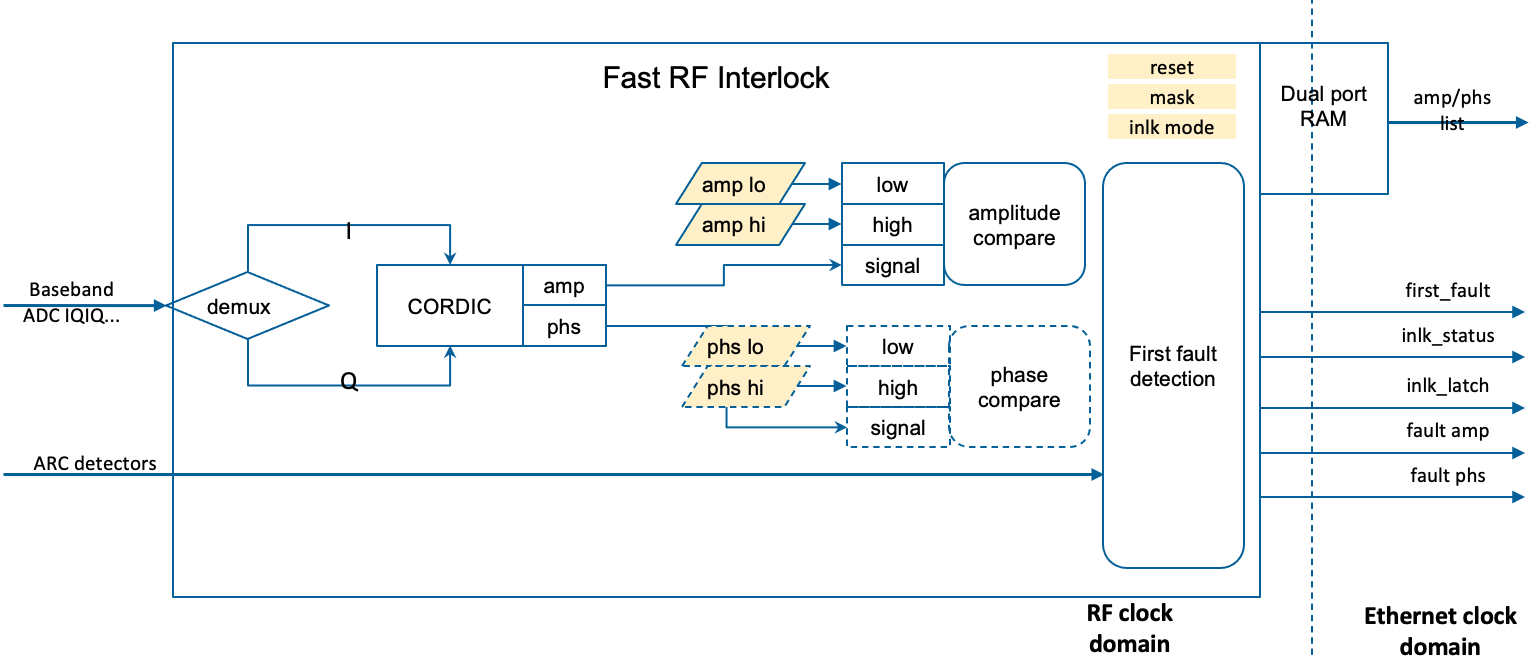}
    \caption{Multichannel RF waveforms and configurable interlock}
    \label{fig:firmware_waveform}
\end{figure}

The total latency of the feedback DSP core is 50 clock cycles or 436 ns.
In parallel, all signals are passed to a multichannel RF waveform buffer with
configuration decimation and cascaded integral-comb (CIC) filter, where a fixed
scaling waveform branches off for realtime RF power interlocking and machine protection purpose,
as shown in Fig.~\ref{fig:firmware_waveform}.
The fast RF power interlock has a configurable mode of comparison between the measured RF amplitude
and a pair of programmable thresholds for disabling the DAC output in case of unexpected RF power or phase measurements,
within $< 1 \mu$s limit to protect down stream high power amplifiers and equipments.

\begin{figure}
    \includegraphics[width=0.9\linewidth]{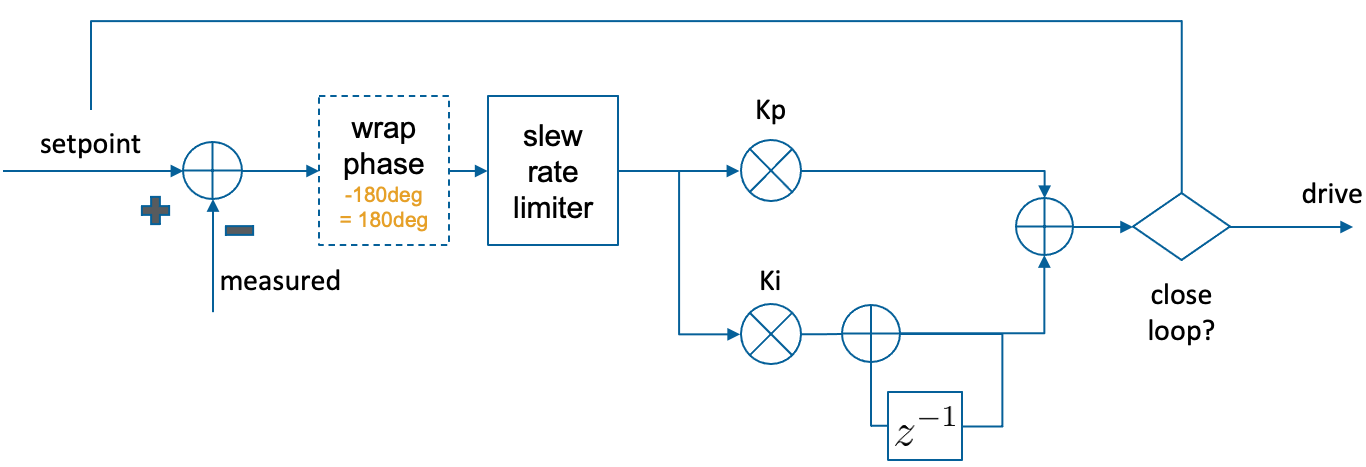}
    \caption{PI controller}
    \label{fig:firmware_pid}
\end{figure}
There are two Proportional--Integral (PI) controllers that are responsible for the feedback loop control
of cavity probe signal's amplitude and phase respectively. 
A classical, scalar PI controller with slew rate limiter and phase wrapping capability is implemented
as shown in Fig.~\ref{fig:firmware_pid} with simple transfer function:

\[
    C(z) = K_p + K_i \frac{1}{1-z^{-1}}, \qquad T=\frac{1}{f_\text{clk}}
\]

We also implemented a state-machine that can synchronously rotate the phase loop
setpoints of both LLRF stations, so that the relative RF bucket can be aligned
between AR and SR. This is required for beam swapping operation for ALS-U.
This \emph{phase ramping} procedure is repeatably
triggered by a common timing event with delay configuration and fault handling.
\begin{figure}[H]
    \includegraphics[width=0.9\linewidth]{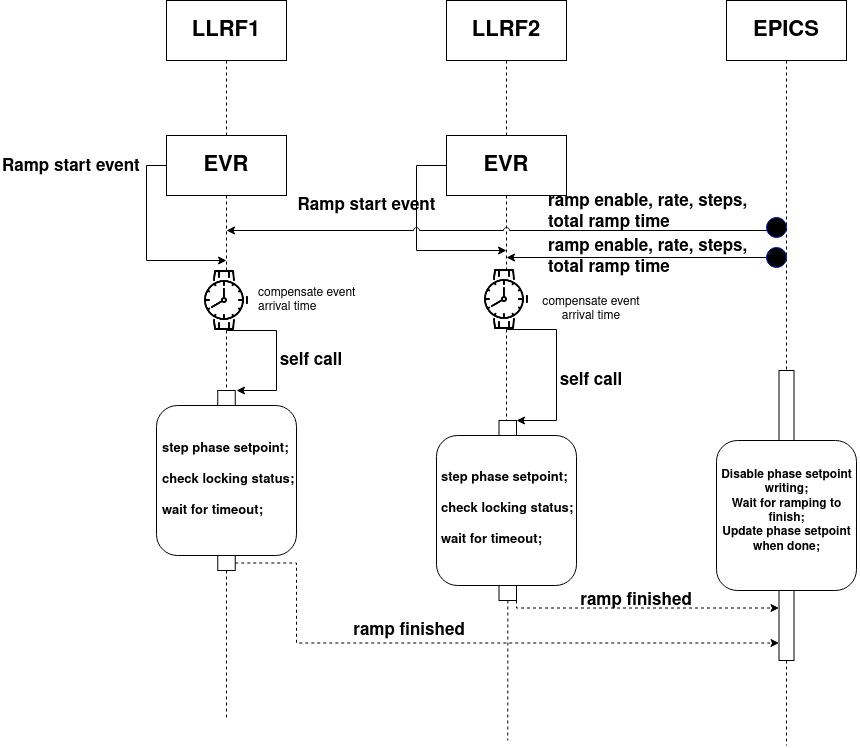}
    \caption{Bucket alignment scheme between AR and SR}
    \label{fig:phase_ramp}
\end{figure}

\section{Software}

Both LLRF stations are managed by a center master interlock PLC through RS485 / Modbus RTU links.
The PLC continuously polling registers through the interrupt service function of the PicoRV32 CPU,
which handles Modbus RTU protocol and retrieves realtime register information for monitoring.

\begin{figure}[H]
    \centering
    \includegraphics[width=0.8\linewidth]{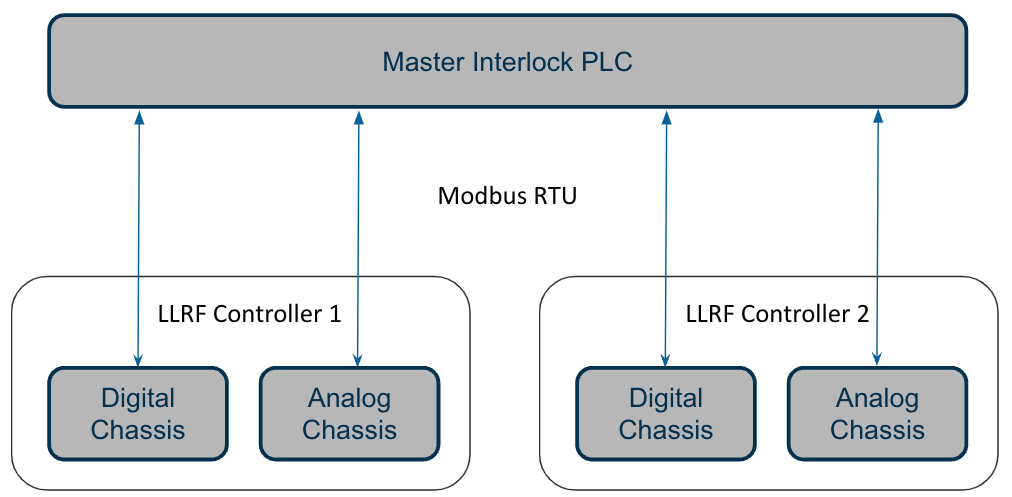}
    \caption{PLC--FPGA interfaces}
\end{figure}

PLC screens are developed for configuration of operation parameters including RF
power / ARC interlock mode and thresholds.

\begin{figure}[H]
    \centering
    \includegraphics[width=0.9\linewidth]{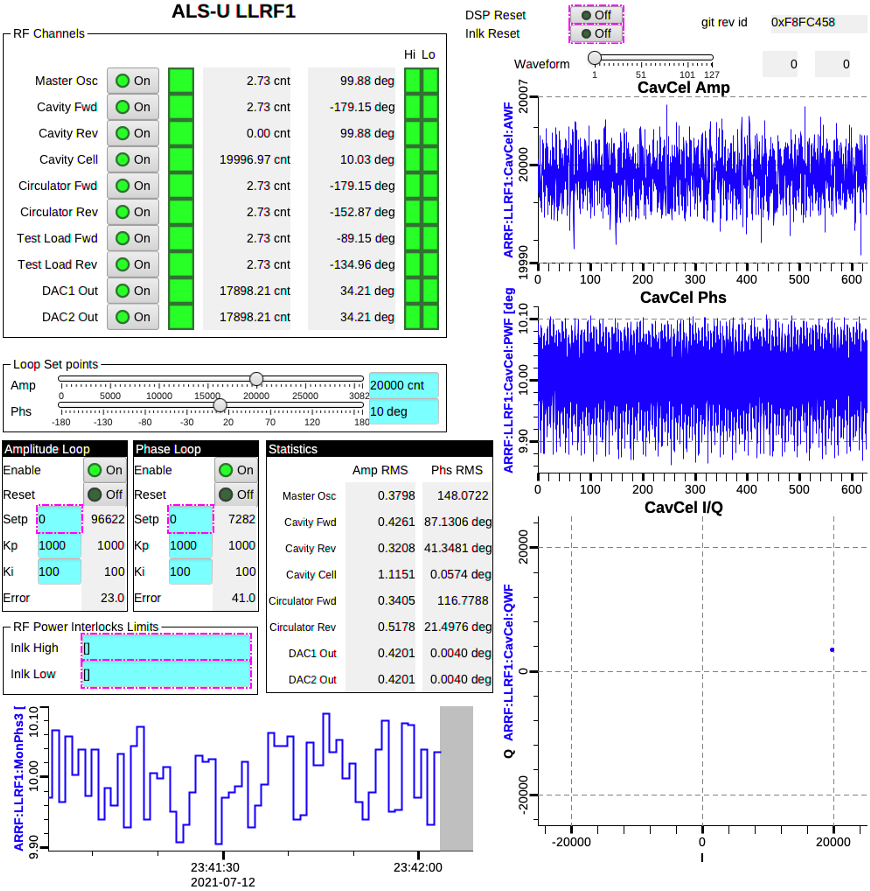}
    \caption{Example Phoebus GUI}
    \label{fig:phoebus}
\end{figure}

An EPICS IOC is built based on single-source-of-truth register mapping, sharing the same architecture with LCLS-II LLRF project.
An example of a working Phoebus engineering screen is shown in Fig.~\ref{fig:phoebus}

\section{Performance}

The single side band $f_\text{LO}$ generation has $> 90$ dB spurious free dynamic range (SFDR), with $-84.6$ dB
MO carrier feedthrough. The measured additive phase noise of LO signal is 29\,fs [1Hz, 1MHz].
The measured RF drive signal also has $> 95$ dB SFDR, with RMS phase noise of 30\,fs [1Hz, 1MHz].

\begin{table}[H]
    \centering
    \begin{tabular}{l|r|c}
    \toprule
        LO Spurious Free Dynamic Range  & $> 90$ & dB     \\
        \midrule
        Carrier feedthrough in LO signal & -84.6 & dB      \\
        \midrule
        RMS Additive LO phase jitter   [1Hz, 1MHz]   & 29 & fs  \\
    \bottomrule
    \end{tabular}
    \caption{Single Side Band LO generation benchmark}
\end{table}

\begin{table}[H]
    \centering
    \begin{tabular}{l|r|c}
    \toprule
        TX Supurious Free Dynamic Range  & $> 95$ & dB     \\
        \midrule
        RMS Additive phase jitter   [1Hz, 1MHz]   & 30 & fs  \\
    \bottomrule
    \end{tabular}
    \caption{RF drive signal (after up-conversion) benchmark}
\end{table}

\begin{figure}[H]
    \includegraphics[width=0.52\linewidth]{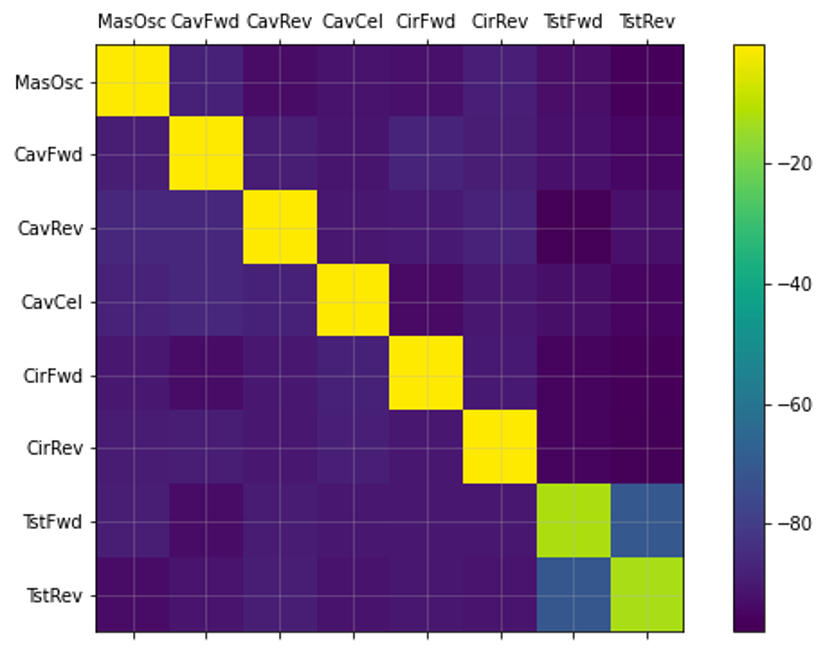}            
    \includegraphics[width=0.42\linewidth]{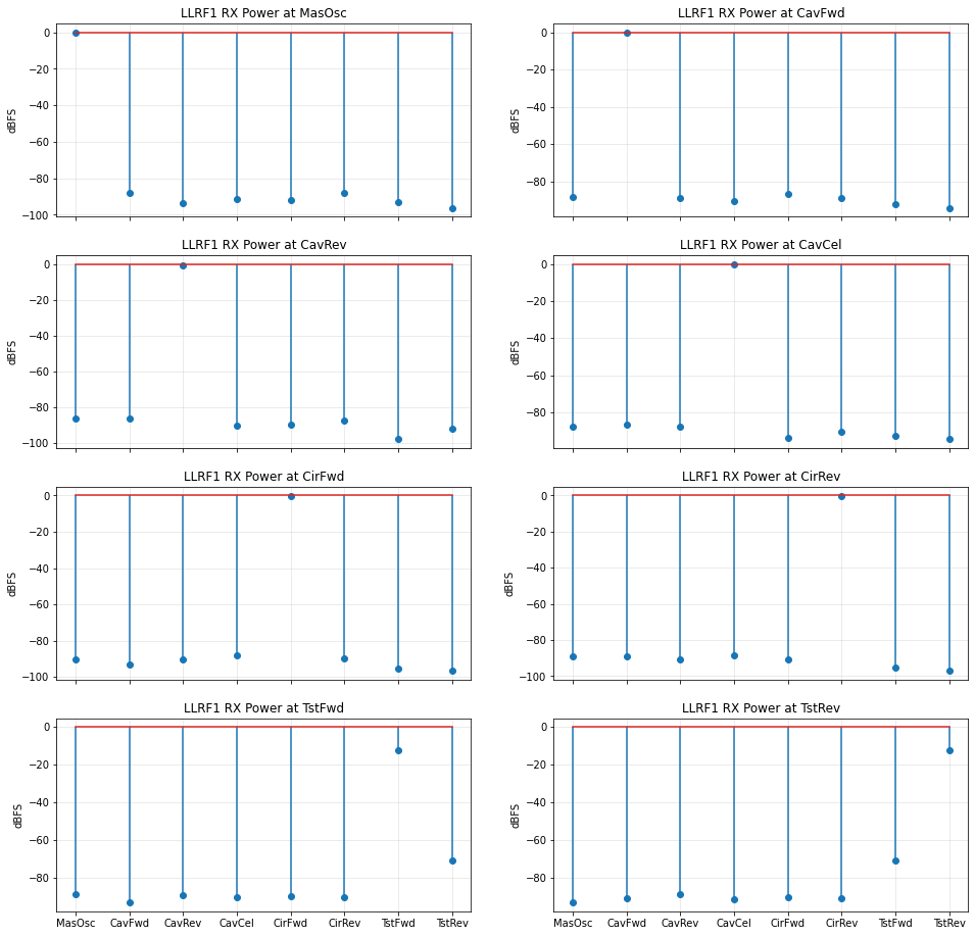}
    \caption{Receiver channel-to-channel isolation}
    \label{fig:isolation}
\end{figure}

The measured receiver channel-to-channel isolation is $> 86$ dB as shown in Fig.~\ref{fig:isolation}.

\begin{figure}[H]
    \centering
    \includegraphics[width=0.45\linewidth]{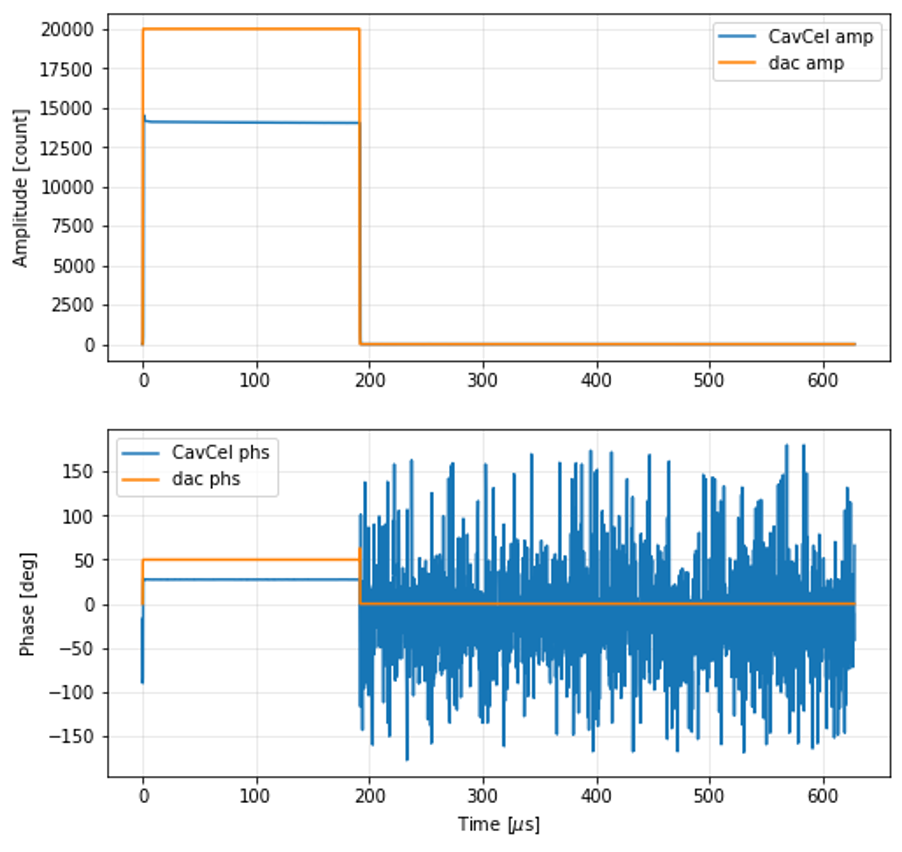}
    \includegraphics[width=0.45\linewidth]{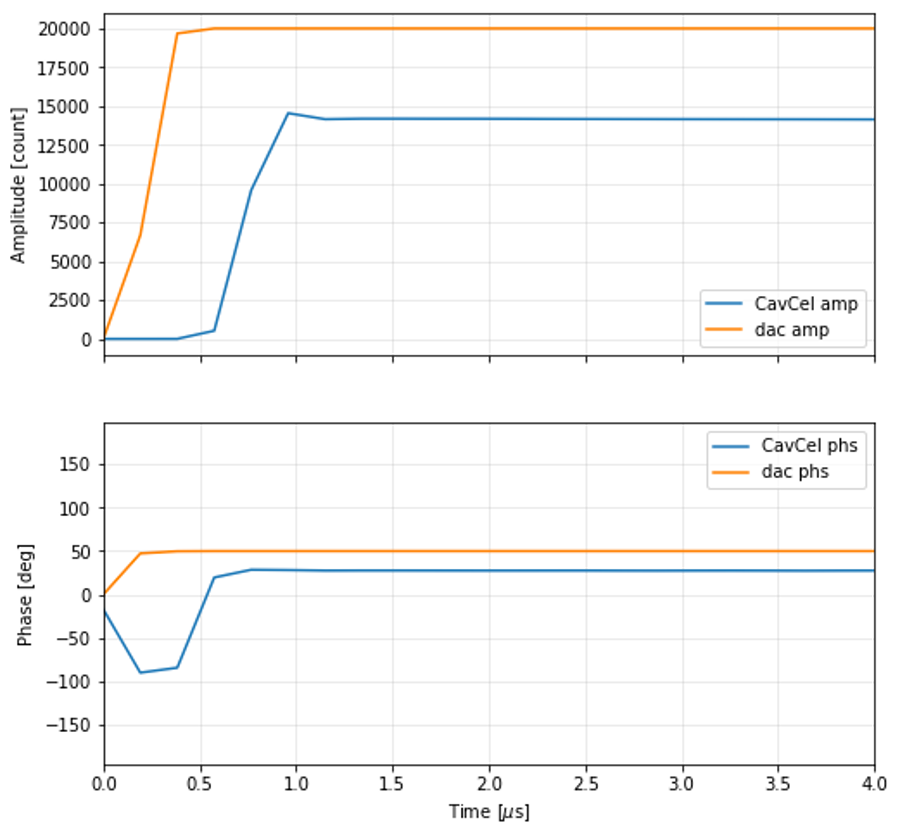}
    \caption{Group Delay measurement}
    \label{fig:group_delay}
\end{figure}

The measured RF-to-RF group delay is $<1 \mu$ second by directly looping back RF
drive signal and one of the receiving signals, when pulsing the drive signal,
as shown in Fig.~\ref{fig:group_delay}.

The feedback loop noise floor is measured also by looping back TX and RX (cavity probe signal),
and using the spectrum analysis of the waveforms. We used an external RF signal analyzzer (R\&S FSWP)
and measured that the out-of-the-loop RMS amplitude stability is $<0.005\%$, and RMS phase loop stability
is $< 0.008 ^\circ$.
\begin{figure}[H]
    \centering
    \includegraphics[width=0.45\linewidth]{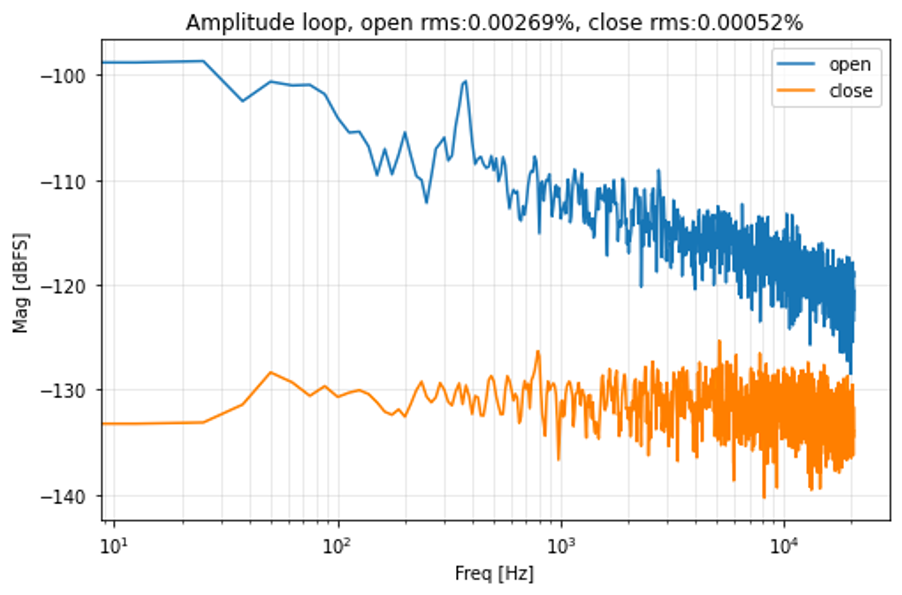}
    \includegraphics[width=0.45\linewidth]{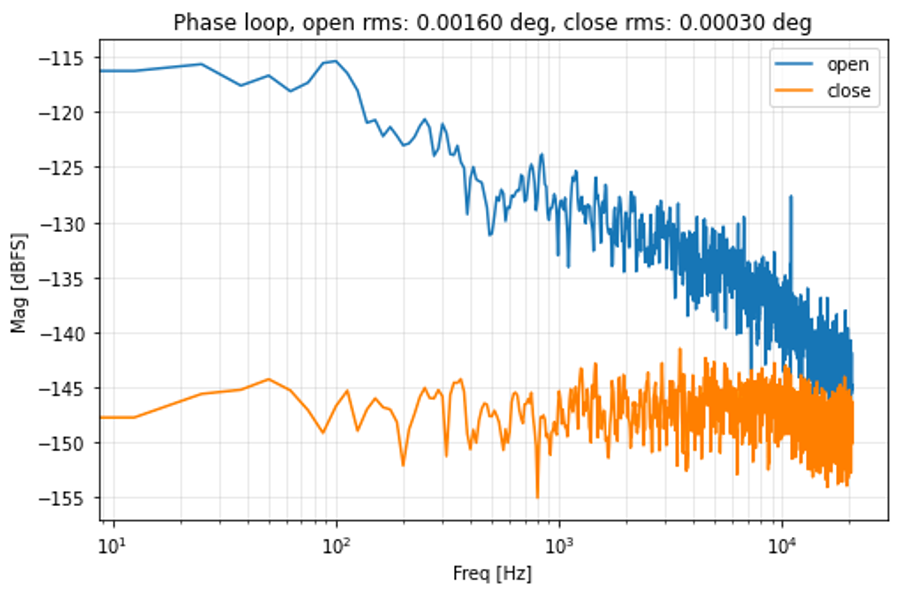}
    \caption{In-loop stability measurement}
    \label{fig:loop_floor_in}
\end{figure}
\begin{figure}[H]
    \centering
    \includegraphics[width=0.45\linewidth]{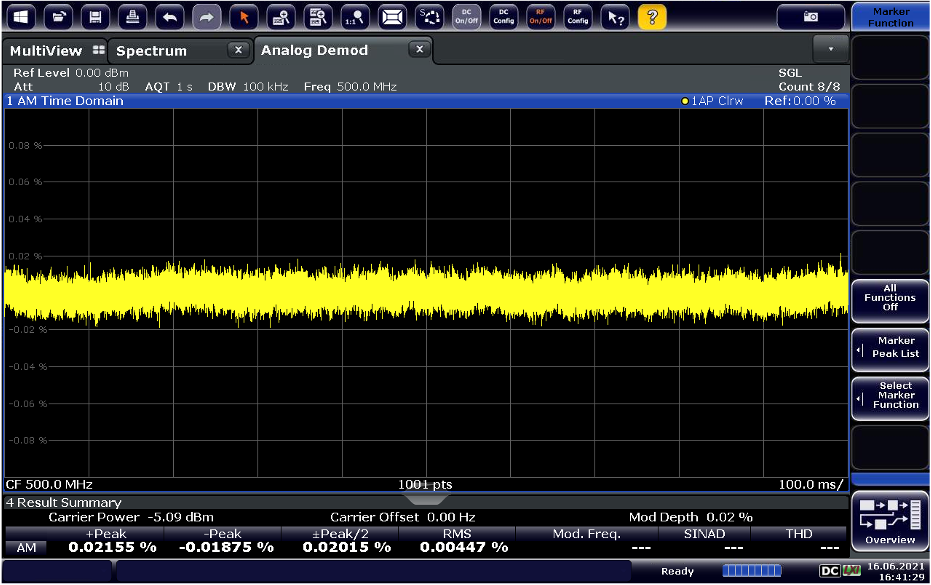}
    \includegraphics[width=0.45\linewidth]{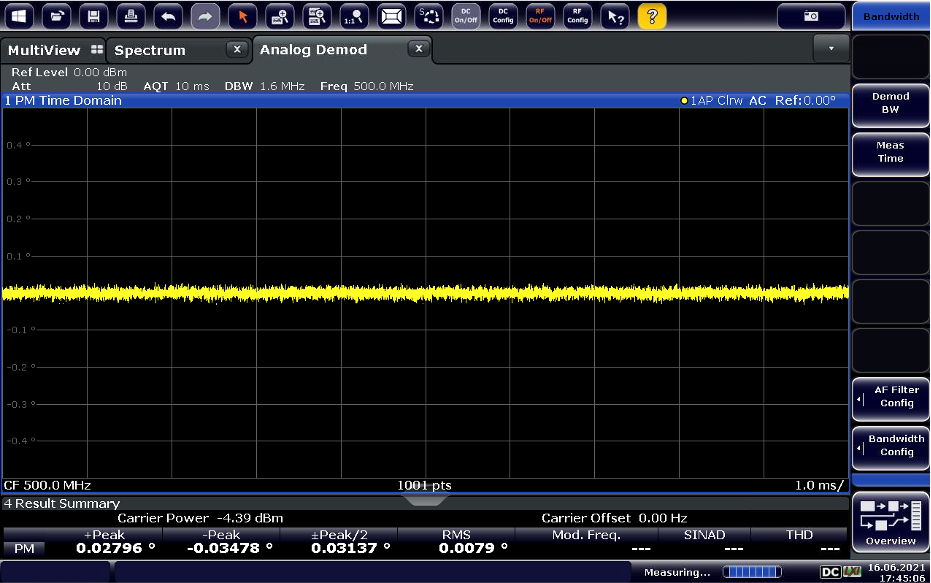}
    \caption{Out-of-loop stability measurement}
    \label{fig:loop_floor_out}
\end{figure}

\begin{figure}[H]
    \centering
    \includegraphics[width=0.45\linewidth]{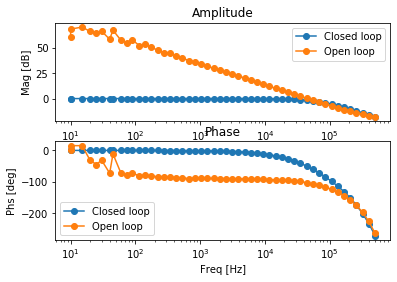}
    \includegraphics[width=0.45\linewidth]{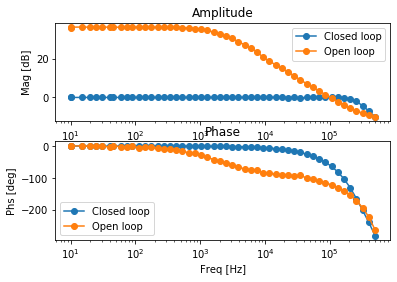}
    \caption{Bode plot of both loops when directly looped back}
    \label{fig:bode}
\end{figure}

We also measured the frequency response of both loops when directly looped back,
by adding a known excitation frequency on the setpoint of either amplitude or
phase loop, and analysis the response spectrum. The resulting Bode plot
is shown in Fig.~\ref{fig:bode}.

\begin{figure}[H]
    \centering
    \includegraphics[width=0.45\linewidth]{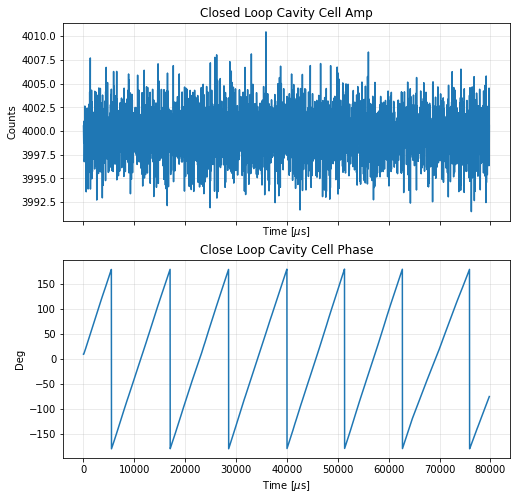}
    \includegraphics[width=0.45\linewidth]{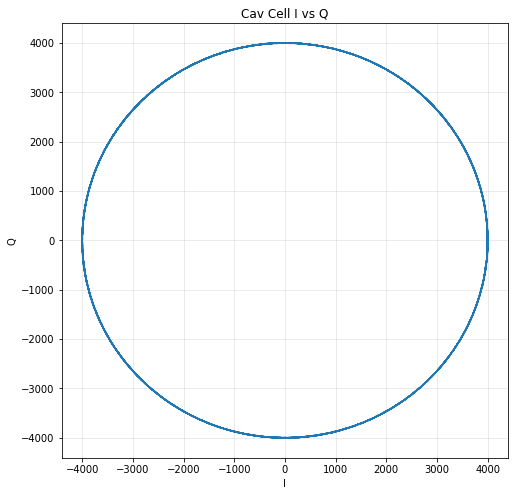}
    \caption{Phase ramping process when both loops are closed}
    \label{fig:phase_ramp_plot}
\end{figure}

The \emph{phase ramping} bucket alignment test result is shown in Fig.~\ref{fig:phase_ramp_plot}.
This is done by ramping of phase setpoint from A to B, at a configurable speed,
when the loop is closed. The ramping successfully executed greater than 7 period of RF cycle,
after being triggered by an timing event, which is required by beam swapping operation between AR and SR.

\section{Conclusion}
\label{sec:conclusion}
The ALS-U AR LLRF system is developed and tested on the bench, where its
performance meats the requirements.
The design is conducted using fully open source hardware, firmware and software
libraries offered by BerkeleyLab.
The same design is also tested in Brazilian Light Source (SIRIUS)
at a different frequency configuration, where
$f_\text{LO} = \frac{23}{34} f_\text{MO} = 479.17\,\text{MHz}$ and
$\frac{f_\text{IF}}{f_\text{Sample}} = \frac{4}{23}$.
The system is ready for cavity test and commissioning at ALS-U project.

\section*{Acknowledgment}

This work is supported by the Office of Science, Office of Basic Energy Sciences, of the U.S.
Department of Energy under Contract No. DE-AC02-05CH11231.

\bibliographystyle{IEEEtran}
\bibliography{IEEEabrv,reference}

\end{document}